# 3 kV Monolithic Bidirectional GaN HEMT on Sapphire

Md Tahmidul Alam[1], Swarnav Mukhopadhyay[1], Md Mobinul Haque[1], Shubhra S. Pasayat[1], and Chirag Gupta[1]
[1]Department of Electrical and Computer Engineering, University of Wisconsin-Madison, WI 53706, USA

*Abstract—* More than 3 kV breakdown voltage was demonstrated in monolithic bidirectional GaN HEMTs for the first time having potential applications in 1200V or 1700V-class novel power converters. The on resistance of the fabricated transistors was ~20 Ω.mm or ~11 mΩ.cm$^2$. Breakdown voltage was optimized by utilizing two field plates in either side of the transistor and optimizing their geometry. Shorter first field plate lengths (≤ 2μm) resulted in higher breakdown voltage and the possible reason for this was discussed. The transistors had a steep subthreshold swing of 92 mV/ dec. The on/off ratio was >10$^5$ and it was limited by the tool capacity. The fabricated 3 kV transistor was benchmarked against the state-of-the-art monolithic bidirectional GaN HEMTs in the performance matrices of breakdown voltage – on resistance, that showed crucial progress.

## I. Introduction

Efficient and reliable extraction of renewable energy is essential to encounter the anticipated global energy shortage due to the depletion of fossil fuels in a few decades [1]-[3]. However, the extraction, storage and conversion of energy from renewable sources is still very inefficient compared to theorical limits because of the lack of high-power, efficient and reliable power converters. Some novel power converters with high power density require bidirectional current and bidirectional blocking capability. Matrix converters, multi-level T-type inverter, current source inverter, solid-state circuit breaker etc. are such examples [4]-[8]. Typically, bidirectional functionality is achieved by connecting two unidirectional transistors in anti-series or anti-parallel configuration [9]-[12]. However, these implementations suffer from high on-resistance, high complexity, low reliability and high form-factor due to high device count (four) and internal contacts. Monolithic bidirectional GaN Transistors / Switches (MBDS) can achieve bidirectional current or blocking capability with a single device hence can potentially mitigate these challenges [13]-[16].

There are several reports on the structure and operation [17],[18], gate-control schemes [19],[20] and substrate termination [21] of monolithic bidirectional GaN HEMTs.

The authors gratefully acknowledge the support of this research from NSF ASCENT (award number ECCS 2328137).
M.T. Alam, S. Mukhopadhyay, M. Haque, S.S. Pasayat, and C. Gupta are with the Department of Electrical and Computer Engineering, University Wisconsin-Madison at Madison, WI 53706, USA (e-mail: malam9@wisc.edu).

One of our previous works demonstrated 1360V GaN MBDS' with a qualitative design guide for breakdown voltage optimization with field plates [22]. However, there is no demonstration of >2 kV breakdown voltage GaN MBDS- which is essential to make 1200 V-class and 1700 V-class power converters. In this work, we report 3 kV (measurement limit of the tool = 3 kV) GaN MBDS for the first time with low on-resistance ($R_{ON}$) of ~20 Ω. mm (11 mΩ cm$^2$) on sapphire substrate. We utilized and optimized two field plates to maximize the breakdown voltage by electric field management. This work has been benchmarked against the state-of-the art monolithic bidirectional GaN HEMTs, indicating crucial advance.

Conventionally, normally-off transistors are preferred than normally-on transistors in power converters due to the ability of normally-off transistors to withstand any accidental damage of the gate driving circuitry. Moreover, normally-off transistors need simpler gate drivers [23], [24]. However, both normally-on and normally off transistors have been commercialized for power electronic applications. Normally-on transistors can be converted to normally-off by cascoding a low-voltage silicon transistor [15],[16], [25]-[28]. In this work, we fabricated normally-on transistors however similar concepts or designs can be applied to normally-off transistors as well for fabrication of high-voltage applications.

## II. Transistor Fabrication and Measurement

The epitaxial structure- 3nm GaN (cap)/ 20nm Al$_{0.24}$Ga$_{0.76}$N (barrier) / 0.7nm AlN/ 1μm UID GaN (channel)/ 2μm semi-insulating GaN (Fe doped, $N_A$ = ~5 × 10$^{18}$ cm$^{-3}$)/sapphire substrate was grown in MOCVD (Fig. 1). Sapphire was chosen over silicon as the substrate material to allow higher breakdown voltage beyond 2 kV [29]-[32]. The fabrication process started with standard solvent cleaning, subsequent ohmic lithography and metal deposition- Ti/Al/Ni/Au (20/120/30/50) nm. The ohmic metal stacks were then annealed at 900°C for 45 seconds in N$_2$ environment. After that, a 750nm deep mesa etch was performed to isolate the devices. Afterwards, 200nm thick Ni gates were deposited in a two-phase deposition- in each phase the plane of the sample was inclined at 30° from the horizontal plane to ensure metal coverage in the sidewall. Then, the surface was passivated by a 320nm thick PECVD Si$_3$N$_4$ layer. Next, two field plate trenches were etched in the Si$_3$N$_4$ layer such that the first and second trench had ~100nm and ~250nm thick Si$_3$N$_4$ left from the AlGaN barrier. Following this, Ni/Au (200/200) nm was deposited as field plate metals on the trenches. Each field plate was connected to the nearest source/ohmic electrode. The field plate lengths ($L_{FP1}$ and $L_{FP2}$) were varied between 1μm to 3μm to optimize the breakdown voltage. Gate length ($L_G$), gate-drain distance ($L_{GD}$), gate-source distance ($L_{GS}$) and width (distance between Mesa edges, Fig. 1(b)) were 2μm, 40μm, 2μm and 100μm respectively. Table-I contains the detailed structural dimensions of the fabricated transistors.

The B1505A (Keysight) source-measurement unit (SMU) was used to perform DC IV, pulsed IV and breakdown voltage measurements. During breakdown measurements the second gate ($G_2$) was shorted to the "drain" ($S_2$) and the devices were immersed in Fluorinert FC-40 to ensure air does not breakdown before transistor breakdown.

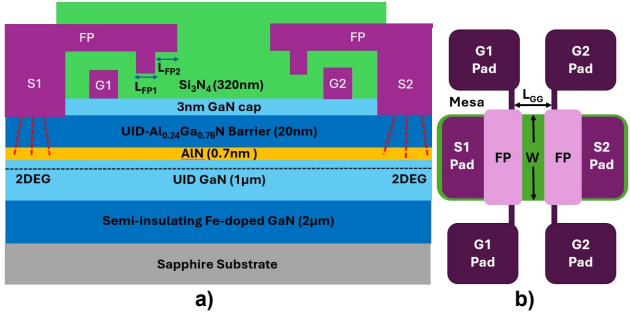

**Fig.1.** Structure of the Monolithic Bidirectional GaN HEMT. a) Cross-section b) Top view. The source pads were 1μm inside the Mesa edge, the field plates covered 2μm beyond the Mesa edge.

TABLE I

| Parameter | Description | Value |
| --- | --- | --- |
| $L_G$ | Gate length | 2μm |
| $L_{GS}$ | Gate to source distance | 2μm |
| $L_{GG}$ | Gate to gate distance | 40μm |
| $L_{FP1}$ | First field plate length | 1μm-3μm |
| $L_{FP2}$ | Second field plate length | 1μm-3μm |
| $L_{GF}$ | Gate to field plate distance | 1μm |
| W | Width | 100μm |
| $T_{FP1}$ | Dielectric under first field plate | 100 nm |
| $T_{FP2}$ | Dielectric under second field plate | 250nm |

### III. RESULTS AND DISCUSSION

#### A. IV Characteristics

The bidirectional IV characteristics of the MBDS are shown in Fig. 2. For $L_{GG}$ = 40μm, the on-resistance was ~20 Ω.mm resulting in a specific resistivity of ~11 mΩ.cm$^2$. The specific resistivity was found by multiplying the on resistance ($R_{ON}$) with the total pitch of the channel ($L_{SD} + 2L_T$). TLM (Transfer Length Measurements) was performed to extract the transfer length ($2L_T$ = ~7μm), contact resistance ($R_c$ = ~0.9 Ω.mm) and sheet resistance ($R_{sheet}$ ~350 Ω/□) of the 2DEG. The extracted value of $R_{ON}$ from IV curves matched closely with the expected $R_{ON}$ from $R_c$ and $R_{sheet}$. The sheet charge density was 8.35 × 10$^{12}$ cm$^{-2}$, and electron mobility was 2010 cm$^2$/V.s, determined by hall measurements.

The threshold voltage ($V_{TH}$) was stable -3.25 V in multiple measurements (assuming 1mA/mm to be the cut-off current) as observed from the transfer characteristics in Fig. 3. The subthreshold swing (SS) was steep 92 mV/dec (between 10$^{-4}$ A/mm to 10$^{-5}$ A.mm), on/off ratio was >10$^5$ and was limited by the tool noise current in the low-current domain. The stable $V_{TH}$, low SS along with high on/off ratio makes this device suitable for high-frequency operations with low conduction and switching losses.

#### B. Breakdown Voltage

Fig. 4 depicts the breakdown response of a transistor with 3 kV breakdown voltage. Both the gate-leakage and drain-leakage current was noticeably stable ($I_D$ = ~90 μA/mm and $I_G$ = ~ 2μA/mm) and bellow 1 mA/mm (breakdown limit) up to the tool limit of 3 kV [33]-[35]. The breakdown voltage was 3 kV (tool limit) for most devices with the first field plate length ($L_{FP}$1) of ≤ 2 μm. However, transistors $L_{FP}$1 > 2 μm exhibited a tendency to have lower breakdown voltage (< 3 kV) as shown in Fig. 5. A possible reason for this trend is that longer total field plate length causes the electric field under the field plates to become stronger. Thus, it results in a high impact ionization rate and causes early transistor breakdown. The detailed mechanism of breakdown is described and justified with TCAD simulations in one of our earlier works [18]. Fig. 6 demonstrates the promising stand of our fabricated MBDS' in the breakdown voltage – on-resistance benchmark compared to the state-of-the-art MBDS'. The reasonable stability of the leakage current and the superior stand in the breakdown voltage – on-resistance benchmark of our fabricated transistors makes them an attractive choice for potential 1700V-class or 1200V-class applications. However, the breakdown field (~75 V/μm) is still much lower than the theoretical critical filed (330 V/μm) of GaN. The critical field or breakdown voltage of the transistors may further be increased by optimizing the field plate number and geometry. We are currently working on this optimization and our future works are expected to publish the relevant studies.

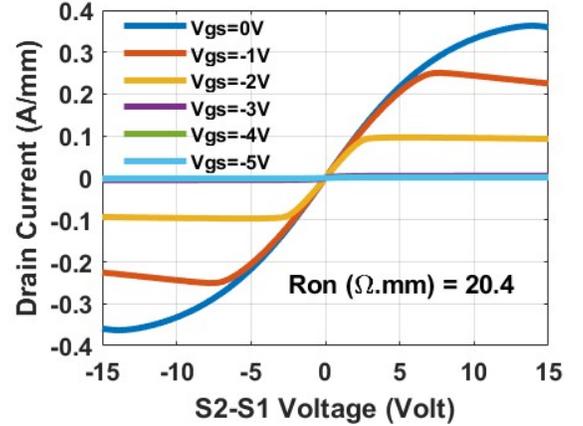

**Fig. 2.** Bidirectional IV characteristics of the MBDS with $L_{GG}$ =40μm, $L_G$ = 2μm, $L_{GS}$ = 2μm.

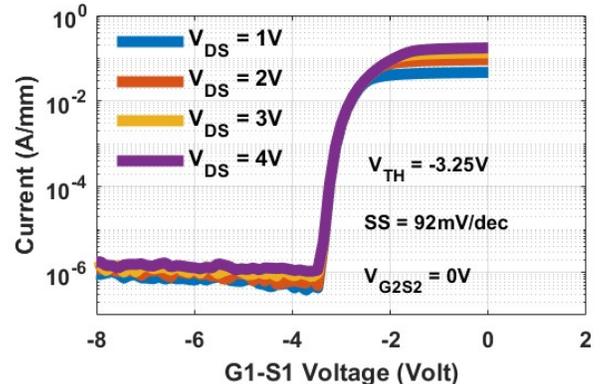

**Fig.3**. Transfer characteristics of the MBDS showing a steep SS of 92 mV/dec and on/off ratio ~ 10$^5$.

*C. Preliminary Dynamic Response*

Pulsed IV measurements with 40V off-state switching voltage (limited by tool capacity) was performed, the dynamic $R_{ON}$ was <10% higher than DC $R_{ON}$ at $V_{DS}$ = 1V (Fig. 7) with 100μs pulse width. The amount of current collapse was less than 10% compared to DC measurements. In these measurements the second gate ($G_2$) and "drain" ($S_2$) were shorted together. Even though the preliminary switching results look promising, the off-state switching voltage should be close to the voltage rating of the application class (~1200V or ~1700V) with <10μs pulse width for practical implementations. The 40 V applied bias in this study was due to the tool limit. Our future studies will focus on high voltage switching tests with appropriate tool setup.

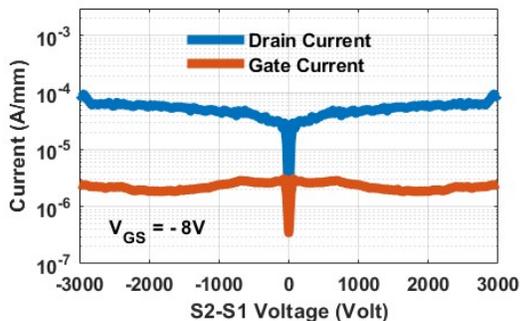

**Fig. 4.** Breakdown response of an MBDS with field plate 1 and 2 length of 1μm and 1.5μm respectively.

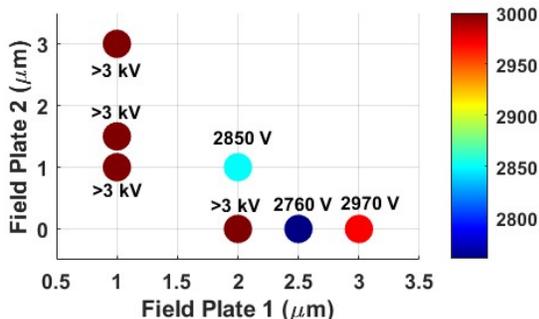

**Fig. 5.** Breakdown voltage variation with field plate dimensions, transistors with smaller first field plate length had higher breakdown voltage.

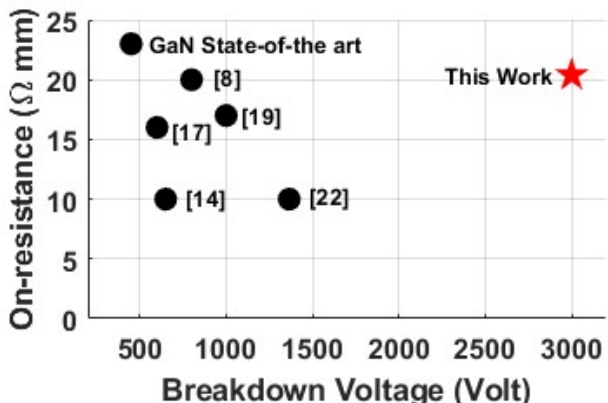

**Fig. 6.** Standing of our fabricated MBDS with state-of-the-art in breakdown voltage- $R_{ON}$ benchmark.

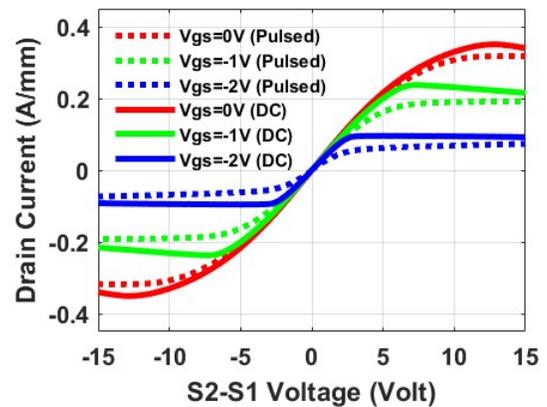

**Fig. 7.** Preliminary dynamic characterization with $V_{DSQ}$ = 40V, $V_{GSQ}$ = -12V, pulse width = 100μs.

## IV. CONCLUSION

Monolithic bidirectional GaN HEMTs with greater than 3 kV (limited by tool capacity) breakdown voltage was demonstrated for the first time for potential applications in 1200V-class or 1700V-class power converters. The breakdown voltage was optimized by utilizing two field plates with varying lengths. The first field plate lengths of ≤ 2μm resulted in higher breakdown voltage, the possible physics behind this was explained. For 3 kV breakdown voltage the on-resistance was low, ~20 Ω.mm (11 mΩ. cm$^2$). In breakdown voltage – on resistance benchmark against the state-of-the-art monolithic bidirectional GaN HEMTs, this work shows significant progress.